\documentclass{article}
\usepackage{amsmath,amssymb,enumerate,float,epsfig,graphicx}
\DeclareGraphicsExtensions{.pdf,.eps}
\renewcommand{\theequation}{\arabic{section}.\arabic{equation}}

\title{ Relativistic modelling of stable anisotropic super-dense star }
\author{ S.K. Maurya \\ Department of Mathematical \& Physical Sciences,\\
College of Arts \& Science, University of Nizwa,\\ Nizwa- Sultanate of Oman \\
e-mail: sunilkumarmaurya1@gmail.com, sunil@unizwa.edu.om \\[2ex]
         Y.K. Gupta
                      \\ Department of Mathematics,\\ Jaypee Institute of Information Technology University,\\
Sector-128 Noida (U.P.), India \\ e-mail: kumar001947@gmail.com\\[2ex]
         M.K. Jasim
                      \\ Department of Mathematical \& Physical Sciences,\\ College of Arts \& Sciences, University of Nizwa,\\ Nizwa, Oman \\
                      e-mail: mahmoodkhalid@unizwa.edu.om}

\begin{document}

\maketitle

\begin{abstract}
In the present article we have obtained new set of exact solutions of
Einstein field equations for anisotropic fluid spheres by using
the Herrera et al.\ \cite{1} algorithm. The anisotropic fluid
solutions so obtained join continuously to Schwarzschild exterior
solution across the pressure free boundary.It is observed that
most of the new anisotropic solutions are well behaved and
utilized to construct the super-dense star models such as neutron
star and pulsars.

\end{abstract}


\section{Introduction}\label{s1}

The first ever exact solution of Einstein's field equation for a
compact object in static equilibrium was obtained by Schwarzschild
in 1916. The static isotropic and anisotropic exact solutions
describing stellar-type configurations have continuously attracted
the interest of physicists Herrera et al.\ \cite{1,41}. Tolman
\cite{2} has proposed an easy way to solve Einstein's field
equations by introducing an additional equation necessary to give
a determinate problem in the form of some ad hoc relation between
the components of metric tensor. According to this methodology,
Tolman \cite{2} has obtained eight solutions of the field
equations, and his important approach still continues in obtaining
the exact interior solutions of the gravitational field equations
for fluid spheres. Buchdahl \cite{3} proposed a famous bound on
the mass radius ratio of relativistic fluid spheres is $2GM/c^{2}r
\le 8/9$, which is an important contribution in order to study the
stability of the fluid spheres. Also Ivanov \cite{4} has given the
upper bound of the red shift for realistic anisotropic star models
which cannot be exceed the values $3.842$ provided the tangential
pressure satisfies a strong energy condition $(\rho \ge p_{r}+
2p_{t})$ and when the tangential pressure satisfies the dominated
energy condition $(\rho \ge p_{t})$. Buchdahl \cite{3} has also
obtained a non-singular exact solution by choosing a particular
choice of the mean density inside the star.

The theoretical investigations of realistic fluid models indicate
that stellar matter may be anisotropic at least in certain density
ranges $(\rho >10^{15}$ gm/cm$^{3})$ (Ruderman \cite{5} and Canuto
\cite{6}) and radial pressure may not be equal to the tangential
pressure of stellar structure. The existence of a solid core due
to presence of the anisotropy in the pressure was thought of by
type-3A super-fluid (Kippenhahm and Weigert \cite{7}), different
form of phase transitions (Sokolov \cite{8}) or by others physical
phenomena. On the scale of galaxies, Binney and Tremaine \cite{9}
have considered anisotropies in spherical galaxies, from a purely
Newtonian point of view. The mixture of two gases (e.g. ionized
hydrogen and electrons or monatomic hydrogen) can be described
formally as an anisotropic fluid (Letelier \cite{10} and Bayin
\cite{11}).The importance of equations of state for relativistic
anisotropic fluid spheres have been investigated by generalizing
the equation of hydrostatic equilibrium to include the effects of
anisotropy (Bowers and Liang \cite{12}). Their study shows that
anisotropy may have non-negligible effects on parameters such as
maximum equilibrium mass and surface red-shift. The relativistic
anisotropic neutron star models at high densities by means of
several simple assumptions showed that there is no limiting mass
of neutron stars for arbitrary large anisotropy which is studied
by Heintzmann and Hillebrandt \cite{13}. However maximum mass of a
neutron star still lies beyond 3-4 $M_\Theta$. Also the solutions
for an anisotropic fluid sphere with uniform density and variable
density are studied by Maharaj and Maartens \cite{14} and Gokhroo
and Mehra \cite{15}, respectively. Most the astronomical objects
have variable density. Therefore, interior solutions of
anisotropic fluid spheres with variable density are more realistic
physically.

Many workers have obtained different exact solutions for isotropic
and anisotropic fluid spheres in different contexts (Delgaty and
Lake \cite{16}, Dev and Gleiser \cite{17}, Komathiraj and Maharaj
\cite{18}, Thirukkanesh and Ragel \cite{19}, Sunzu et al.\
\cite{20}, Harko and Mak \cite{21}, Mak and Harko \cite{22},
Chaisi and Maharaj \cite{23}, Maurya and Gupta \cite{24,25,26},
Feroze and Siddiqui \cite{27}, Pant et al.\ \cite{28,29,30}, Bhara
et al.\ \cite{31}, Monowar et al.\ \cite{32}, Kalam et al.\
\cite{33}, Consenza et al.\ \cite{34}, Krori \cite{35}, Singh et
al.\ \cite{36}, Patel and Mehta \cite{37}, Malaver \cite{38,39},
Escupli et al.\ \cite{40}, Herrera and Santos \cite{41,42},
Herrera et al.\ \cite{43,44}).

The present paper consists nine sections, Section~\ref{s1} contains
introduction; Section~\ref{s2} contains metric, its components and
the field equations. Section~\ref{s3} embodies the solutions of
anisotropic fluid spheres in different contexts. Section~\ref{s4}
contains the expressions for density and pressure are mentioned
for each fluid sphere.Section~\ref{s5} consists the various
physical conditions to be satisfied by the anisotropic fluid
spheres. The analytical behavior of the solutions under the
physical conditions (mentioned in Section~\ref{s5}) are mentioned
in the section~\ref{s6}. Section~\ref{s7} describes the evaluation of
arbitrary constants involved in the fluid solutions by means of
the smooth joining of Schwarzschild metric at the pressure free
interface $r=a$. The stability of models is proposed in the
section~\ref{s8} and finally section~\ref{s9} includes the
physical analysis of the solutions so obtained along with the
concluding remarks.

\section{Metric, components and Field equations}\label{s2}

The line element of static spherical symmetric space time in the
curvature coordinates $x^{i} =(t, r, \chi, \xi )$ can be furnished as below, 
\begin{equation} \label{eq2.1}
ds^{2} =B^{2} (r)dt^{2} -\psi ^{-1} (r) dr^{2} -r^{2} (d\chi ^{2} +\sin ^{2} \chi d\xi ^{2} )
\end{equation}
Einstein's field equations given as
\begin{equation} \label{eq2.2}
-\kappa T_{i}^{j} =R_{i}^{j} -\frac{1}{2} R\delta _{i}^{j}
\end{equation}
where, $\kappa =\frac{8\pi G}{c^{4} }$.

The components of the energy momentum tensor for spherically
symmetric anisotropic fluid distribution is postulated in the form:
\begin{equation} \label{eq2.3}
T_{i}^{j} =(c^{2} \rho +p_{t} )v_{i} v^{j} -p_{t} \delta _{i}^{j}
+(p_{r} -p_{t} ) \chi _{i} \chi ^{j},
\end{equation}
where $v^{i} $ is four-velocity $B v^{i} =\delta _{0}^{i} $, $\chi
^{i} $ is the unit space like vector in the direction of radial
vector, $\chi ^{i} =\sqrt{B} \delta _{1}^{i} $, $\rho $ is the
energy density, $p_{r} $ is the pressure in direction of $\chi
^{i} $ (normal pressure) and $p_{t} $is the pressure orthogonal to
$\chi _{i} $ (transversal or tangential pressure). Suppose radial
pressure is not equal to the tangential pressure i.e.\ $p_{r} \ne
p_{t} $, otherwise if radial pressure is equal to transverse
pressure i.e.\ $p_{r} =p_{t} $, it corresponds to isotropic or
perfect fluid distribution.\ Let the measure of anisotropy $\Delta
=\kappa (p_{t} -p_{r} )$ ,which is called the anisotropy factor
(Herrera and Ponce de Leon \cite{45}). The term $2(p_{t} -p_{r}
)/r$ appears in the conservation equations $T_{j ; i}^{i} =0$
(where, semi colon denotes the covariant derivative) which is
representing a force due to anisotropic nature of the fluid.\ When
$p_{t} >p_{r} $, then the direction of force to be outward
direction and inward when $p_{t} <p_{r} $.\ However, if $p_{t}
>p_{r} $, then the force allows the construction of more compact
object when using anisotropic fluid than when using isotropic
fluid (Gokhroo and Mehra \cite{15}).

In view of metric \eqref{eq2.1}, the Einstein field equations \eqref{eq2.2} give
\begin{align} \label{eq2.4}
&\left(\frac{8 \pi G}{c^{2} } \rho \right)=\frac{1-\psi }{r^{2} } -\frac{\psi '}{r},\quad \left(\frac{8 \pi G}{c^{4} } p_{r} \right)=\frac{2B'\psi }{B r} +\frac{\psi -1}{r^{2} }\\
&\label{eq2.5}
\psi ' \left(\frac{B'}{B} +\frac{1}{r} \right)+2 \psi \left(\frac{B''}{B} -\frac{B'}{r B} -\frac{1}{r^{2} } \right)=2 \left(\Delta -\frac{1}{r^{2} } \right),
\end{align}
where ``dash'' denote the derivative with respect to $r$.

\section{Classes of solutions}\label{s3}
\setcounter{equation}{0}

The field Eq.\,\eqref{eq2.5} has two dependent variables $B(r)$
and $\psi(r)$, therefore Eq.\,\eqref{eq2.5} can admit infinity
many solutions for different choices of $B(r)$ and $\psi(r)$ but
all these solutions may or may not satisfy the physical
conditions for the fluid spheres. for a given $B(r)$, the
Eq.\,\eqref{eq2.5} reduces to first order ordinary
differential equation in $\psi(r)$.\ For its physically valid
solution we will have to choose the metric potential $B(r)$ such that $B(0)$ is non zero positive finite. This is a sufficient condition
for a static fluid sphere to be regular at the centre.

Let us take $B(r)$ of the form
\begin{align}
B(r)=D(1-c_{0}r^{2})^{n}, \ \ \text{where $(n\ne 0)$, $D$ is positive arbitrary constant}. \label{eq3.1}
\end{align}

Maurya and Gupta \cite{22,23} have already obtained all possible
anisotropic solutions of Einstein field equations for $n$ is
positive integer $\ge 1$ with $c_{0} <0$ and $-1< n < 0$,
$c_{0}>0$ with different anisotropic factor $\Delta $.

But in present problem, we consider $B(r) =D (1-c_{0}r^{2})^{n}$
with $n=-1, -2$ and $-3$ and anisotropy factor
\begin{align}
\Delta =\frac{\Delta _{0} c_{0} (c_{0} r^{2} )^{-n} }{[ 1-(n+1) c_{0} r^{2} ]^{1+n} } ; \ \text{where, $c_{0}$ and $\Delta _{0} $ are positive constants.} \label{eq3.2}
\end{align}
Herrera et al\ \cite{1} have proposed an algorithm for all
possible spherically symmetric anisotropic solutions of Einstein
field equations.

\begin{subequations}
By using the Herrera et al\ \cite{1} algorithm, the equations
\eqref{eq2.5} reduces in the form as:
\begin{align}
\psi '+ \left(\frac{2y'}{y} +2y-\frac{6}{r} +\frac{4}{r^{2} y} \right) \psi =\frac{2}{y} \left(\Delta -\frac{1}{r^{2} } \right) \label{eq3.3a}
\end{align}
and
\begin{align}
y(r)=\frac{B'(r)}{2 B(r)} +\frac{1}{r} \,. \label{eq3.3b}
\end{align}
On integrating \eqref{eq3.3a}, we can obtain $\psi $ as:
\begin{align}
\psi =\frac{\displaystyle r^{6} \left[-2\int \frac{y(r) (1+\Delta (r) r^{2} ) e^{\int [(4/r^{2} y(r))+2y(r)] dr } }{r^{8} } dr+A\right]}{y^{2} (r) e^{\int [(4/r^{2} y(r))+2y(r)] dr } } \label{eq3.3c}
\end{align}
and
\begin{align}
y (r)=\frac{[ 1-(n+1) c_{0} r^{2} ]}{r (1-c_{0} r^{2} )} \,.\label{eq3.3d}
\end{align}
\end{subequations}
which further supply
\begin{subequations}
\begin{align}
\psi =\phi f^{3} e^{2\phi } \left[\Delta _{0} \int \frac{e^{-2\phi } }{f^{2} } d\phi -\int \frac{e^{-2\phi } }{\phi f^{2} } d\phi \right]+A\phi f^{3} e^{2\phi }\ \text{for} \ n=-1\label{eq3.4a}
\end{align}
and
\begin{align}
\psi
&=\frac{\phi f^{2-n} }{g^{2/(n+1)} } \left[\Delta _{0} \int \frac{f^{n-1} }{\phi ^{n+1} g^{n(n+3)/(n+1)} } d\phi -\int \frac{f^{n-1} }{\phi ^{2} g^{(n-1)/(n+1)} } d\phi \right]+A\frac{\phi f^{2-n} }{g^{2/(n+1)} }; \nonumber\\
&\hskip3.9in\text{for} \ \ n\ne -1.\label{eq3.4b}
\end{align}
\end{subequations}
where, $\phi =c_{0} r^{2} $, $f=(1-\phi )$ and $g=[ 1-(n+1) \phi ]$.

The Eq.\,\eqref{eq3.4a} and Eq.\,\eqref{eq3.4b} give the following solutions:
\begin{subequations}
\begin{align}
\psi =A. \phi f^{3} e^{2\phi } +(1-2\phi ) f^{2} -(6+\Delta _{0} ) f^{3} \phi e^{2\phi -2} Ei (2-2\phi )+ \Delta _{0} \phi f^{2} ; \ \text{for} \ n=-1, \label{eq3.5a}
\end{align}
where $Ei (2-2 \phi )=\log (2-2\phi )+\sum \limits_{N=1}^{\infty }\frac{(2-2\phi )^{N} }{N ! N} $
\begin{align}
\psi
&=\left[
A\phi(1+\phi )^{2} f^{4} +\left[f^{2} \frac{ (15\phi ^{4} -25\phi ^{2} +8)}{8} +\frac{15}{16} \phi (1+\phi )^{2} f^{4} \log \left(\frac{f}{1+\phi } \right)\right]\right.\nonumber\\
 & \quad \ \ \left. +\frac{\Delta _{0} }{16} \phi f^{2} \left[2 (\phi ^{2} -\phi +2)+f^{2} (1+\phi )^{2} \log \frac{1-\phi }{1+\phi } \right]\right]; \ \text{for} \ \ n=-2; \label{eq3.5b}\\
\psi
&=\left[A \phi f^{5} (1+2\phi )-\frac{1}{243} f^{5} \left[\frac{3 h}{f^{3} } -320 \phi \log \left(\frac{1+2\phi }{f} \right) (1+2\phi )\right]\right.\nonumber \\ & \quad  \  \ +\Delta _{0} \phi  f^{3} (1+2\phi )(1-3\phi +3\phi ^{2} )\bigg];  \ \text{for} \ n=-3 ; \label{eq3.5c}
\end{align}
\end{subequations}
where $h=(-81+130\phi +36\phi ^{2} -720\phi ^{3} +320\phi ^{4} )$.

\section{Expressions for energy density, pressures for different values of $n$}\label{s4}
\renewcommand{\theequation}{4.\arabic{equation}a}
\setcounter{equation}{0}

\subsection*{(a) For $n=-1$}
\begin{align}
\left(\frac{8 \pi G }{c^{2} } \rho \right) &=c_{0}[A f^{2}
e^{2\phi } (4\phi ^{2} +5\phi -1)+(6 +\Delta _{0} )f^{2} (3-5\phi
-4\phi ^{2} ) e^{2\phi -2} Ei (2-2\phi )\nonumber\\ &\qquad
+(6-11\phi ^{2} +2\phi ^{3} )+ \Delta _{0} [(-3+10 \phi -7\phi
^{2} )+2 \phi e^{2\phi -2} f^{3} \bar{E}i (2-2\phi
)]],\label{eq4.1a}\\ \left(\frac{8 \pi G }{c^{4} } p_{r} \right)
&=c_{0} [ {A f^{2} (1+\phi ) e^{2\phi } +(2-10\phi +7\phi ^{2}
-2\phi ^{3} )-6 f^{2} (1+5\phi ) e^{2\phi -2} Ei (2-2\phi )}
\nonumber\\ &\qquad{+\Delta _{0} (1-\phi ^{2} ) [1-e^{2\phi -2} f
Ei(2-2\phi )]} ],\label{eq4.2a}\\ \left(\frac{8 \pi G }{c^{4} }
p_{t} \right) &=c_{0} [ {A f^{2} (1+\phi ) e^{2\phi } +(2-10\phi
+7\phi ^{2} -2\phi ^{3} )-6 f^{2} (1+5\phi ) e^{2\phi -2} Ei
(2-2\phi )} \nonumber\\ &\qquad {+\Delta _{0} [(1+\phi -\phi ^{2}
)-(1-\phi ^{2} )e^{2\phi -2} f Ei(2-2\phi )]} ].\label{eq4.3a}
\end{align}
where, $\bar{E}i (2-2 \phi )=\frac{-2}{(2-2\phi )} -2\sum \limits_{N=1}^{\infty }\frac{(2-2\phi )^{N-1} }{N! } $.

\subsection*{(b) For $n=-2$}
\renewcommand{\theequation}{4.\arabic{equation}b}
\setcounter{equation}{0}
\begin{align}
\left(\frac{8 \pi G }{c^{2} } \rho \right)
&=c_{0} \bigg[ \frac{48+85\phi -350\phi ^{2} +90\phi ^{3} +330\phi ^{4} -195\phi ^{5} +30 \phi f^{3} (1+\phi )}{8} \nonumber\\
&\qquad -\frac{1}{16} f^{3} \left[ { 45-15}\phi -285\phi ^{2} {-225}\phi ^{3} {-} \Delta _{0 } (15\phi ^{3} +19\phi ^{2} +\phi -3) \right]\log \left(\frac{1-\phi }{1+\phi } \right) \nonumber\\
&\qquad\quad -\frac{\Delta _{0 } f}{8} [11\phi ^{3} -16\phi ^{2} +19\phi -6+2f^{2} \phi (1+\phi )]\nonumber\\
&\qquad\quad +A f^{3} (-3+\phi +19\phi ^{2} +15 \phi ^{3} ) \bigg],\label{eq4.1b}\\
\left(\frac{8 \pi G }{c^{4} } p_{r} \right)
&=c_{0} \left[ {A (1+3\phi ) f^{3} (1+\phi )^{2} +\left[\frac{1}{8} (16-49\phi -50\phi ^{2} +90\phi ^{3} +30\phi ^{4} -45\phi ^{5} )\right]}\right. \nonumber\\
&\qquad\quad+\frac{(15+\Delta _{0} )}{16} f^{3} (1+3\phi ) (1+\phi )^{2} \log \left(\frac{1-\phi }{1+\phi } \right)\nonumber\\
&\qquad\quad\left.+\frac{\Delta _{0} }{16} f (1+3\phi ) (4-2\phi +2\phi ^{2} ) \right], \label{eq4.2b}
\end{align}
\begin{align}
\left(\frac{8 \pi G }{c^{4} } p_{t} \right) &=c_{0} \left[ A
(1+3\phi ) f^{3} (1+\phi )^{2} +\left[\frac{(16-49\phi -50\phi
^{2} +90\phi ^{3} +30\phi ^{4} -45\phi ^{5} )}{8}
\right]\right.\nonumber \\ &\qquad\quad +\frac{(15+\Delta _{0}
)}{16} f^{3} (1+3\phi ) (1+\phi )^{2} \log \left(\frac{1-\phi
}{1+\phi } \right)\nonumber\\ &\qquad\quad \left.+\frac{\Delta
_{0} }{16} { (4}+{6}\phi +2\phi ^{2} +{26}\phi ^{3} {-6}\phi ^{4}
{)} \right]. \label{eq4.3b}
\end{align}

\subsection*{(c) For $n=-3$}
\renewcommand{\theequation}{4.\arabic{equation}c}
\setcounter{equation}{0}

\begin{align}
\left(\frac{8 \pi G }{c^{2} } \rho \right) &=c_{0} \bigg[-A
(3-15\phi +90\phi ^{3} -165\phi ^{4} +117\phi ^{5} -34\phi ^{6} )
\nonumber\\ &\qquad\quad+\frac{1}{81} (21320-26565\phi -55034\phi
^{2} +156004 \phi ^{3} -11040 \phi ^{4} +29120 \phi ^{5}
)\nonumber\\ &\qquad\quad+\frac{320}{243} f^{4} (-1+5 \phi +26
\phi ^{2} ) \log \left(\frac{1+2\phi }{1-\phi } \right)\nonumber\\
&\qquad\quad -\Delta _{0} f^{2} (3-14\phi -10\phi ^{2} +93\phi
^{3} -90\phi ^{4} )\bigg], \label{eq4.1c}\\ \left(\frac{8 \pi G
}{c^{4} } p_{r} \right) &=c_{0} \bigg[A f^{4} (1+7\phi +10\phi
^{2} ) \!-\!(\!-\!356+411\phi +146\phi ^{2} \!-\!3820 \phi ^{3}
\!-\!1680 \phi ^{4} +1600 \phi ^{5} )\nonumber\\
&\qquad\quad+\frac{320}{243} f^{4} (1+2\phi )(1+5\phi )\log
\left(\frac{1+2\phi }{1-\phi } \right)\nonumber\\
&\qquad\quad+\Delta _{0} f^{2} (1+4\phi -8\phi ^{2} -9\phi ^{3}
+30\phi ^{4} ) \bigg], \label{eq4.2c}\\ \left(\frac{8 \pi G
}{c^{4} } p_{t} \right) &=c_{0} \bigg[A f^{4} (1+7\phi +10\phi
^{2} ) +(356\!-\!411\phi \!-\!146\phi ^{2} +3820 \phi ^{3} +1680
\phi ^{4} \!-\!1600 \phi ^{5} )\nonumber \\ &\qquad\quad
+\frac{320}{243} f^{4} (1+2\phi )(1+5\phi )\log
\left(\frac{1+2\phi }{1-\phi } \right)\nonumber\\ &\qquad\quad
+\Delta _{0} \left\{f^{2} (1+4\phi -8\phi ^{2} -9\phi ^{3} +30\phi
^{4} ) +\phi ^{3} (1+2\phi )^{2} \right\} \bigg]. \label{eq4.3c}
\end{align}

\section{Reality and Physical (well behaved) conditions for anisotropic solutions}\label{s5}

The physically meaningful anisotropic solution for the Einstein's
field equations must satisfy some physical conditions (Mak and
Harko \cite{20}, and Maurya and Gupta \cite{22}):
\begin{enumerate}[(i)]
\item The solution should be free from physical and geometrical singularities i.e. pressure and energy density should be finite
at the centre and metric potential $B(r)$ and $\psi(r)$ have non zero positive values.

\item The radial pressure $p_{r }$ must be vanishing but the tangential pressure $p_{t}$ may not vanish
at the boundary $r = r_{a}$ of the fluid sphere. However, the radial pressure is equal to the tangential pressure
at the centre of the fluid sphere.

\item The density $\rho$ and radial pressure $p_{r}$ and tangential pressure $p_{t}$  should be positive inside the star.

\item $(dp_{r}/dr)_{r=0} = 0$ and $(d^{2}p_{r}/dr^{2})_{r=0}<0$ so that the pressure gradient $dp_{r}/dr$ is negative for $0\le r\le r_{a}$.

\item $(dp_{t}/dr)_{r=0}=0$ and $(d^{2}p_{t}/dr^{2})_{r=0}<0$ so that the pressure gradient $dp_{t}/dr$ is negative for $0\le r\le r_{a} $.

\item $(d\rho/dr)_{r=0} = 0$ and $(d^{2}\rho/dr^{2})_{r=0}<0$ so that the density gradient $d\rho/dr$ is negative for $0\le r\le r_{a} $.

Conditions (iv)-(vi) imply that pressure and density should be maximum at the centre and monotonically decreasing towards the surface.

\item Inside fluid ball the speed of sound should be less than that of light i.e.
\[
0\le \sqrt{\frac{dp_{r} }{c^{2} d\rho } } <1, \  0\le
\sqrt{\frac{dp_{t} }{c^{2} d\rho } } <1
\]

In addition the velocity of sound monotonically decreasing away
from the centre, the velocity of sound is increasing with the
increase of density i.e.\ $\frac{d}{dr} \left(\frac{dp_{r} }{d\rho
} \right)<0$ or$\left(\frac{d^{2} p_{r} }{d\rho ^{2} } \right)>0$
and $\frac{d}{dr} \left(\frac{dp_{t} }{d\rho } \right)<0$ or
$\left(\frac{d^{2} p_{t} }{d\rho ^{2} } \right)>0$ for $0\le r\le
r_{a} $.\ In this contexts it is worth mentioning that the
equation of state at ultra-high distribution has the property that
the speed of sound is decreasing outwards (Canuto \cite{30}). 
\item A physically reasonable energy-momentum tensor has to obey the energy conditions $\rho \geq  p_{r}+2p_{t}$
(strong energy condition) and $\rho+ p_{r}+ 2p_{t}\geq 0$.
\item The red shift at center $Z_{0}$ and at surface $Z_{a}$ should be positive, finite and both bounded.
\item The anisotropy factor $\Delta$ should be zero at the center and must be increasing towards the surface.
\end{enumerate}

\section{Physical properties of the new solutions}\label{s6}

\renewcommand{\theequation}{6.\arabic{equation}a}
\setcounter{equation}{0}

\subsection*{(a) For $n=-1$}

The expression for pressures and density at the centre are as:
\begin{align}
&\left(\frac{8 \pi G }{c^{4} } p_{r} \right)_{r=0}
=\left(\frac{8\pi G}{c^{4} } p_{t} \right)_{r=0} =c_{0}
\left(A+2+\frac{\Delta _{0} e^{2} -Ei(2) (6+\Delta _{0} )}{e^{2} }
\right),\label{eq6.1a}\\ &\left(\frac{8 \pi G }{c^{2} } \rho
\right)_{r=0} = c_{0} \left(-A+6+\frac{ Ei(2) (18+3\Delta _{0}
)-3e^{2} \Delta _{0} }{e^{2} } \right).\label{eq6.2a}
\end{align}
The pressure and density should be non zero positive at the centre and consequently $A$ satisfy the following inequality:
\begin{align}
A>\!\left(-2+\frac{Ei(2) (6+\Delta _{0} )-\Delta _{0} e^{2} }{e^{2} } \right) \text{ and } A<\!\left(6+\frac{ Ei(2) (18+3\Delta _{0} )-3\Delta _{0}
 e^{2} }{e^{2} } \right)\!. \label{eq6.3a}
\end{align}
The ratio of pressure-density should be positive and less than $1$
at the centre i.e. $\frac{p_{0} }{\rho _{0} c^{2} } \le 1$ which
gives the following inequality,
\begin{align}
\left(\frac{p_{r} }{\rho c^{2} } \right)_{r=0} =\left(\frac{p_{t}
}{\rho c^{2} } \right)_{r=0} =\left(\frac{e^{2} (A+2)+\Delta _{0}
e^{2} -Ei(2) (6+\Delta _{0} )}{e^{2} (-A+6)+Ei(2) (18+3\Delta _{0}
)-3\Delta _{0} e^{2} } \right)\le 1\label{eq6.4a}
\end{align}
By Eq.\,\eqref{eq6.4a}, we get
\begin{align}
A\le 2+\frac{Ei(2) (6+\Delta _{0} )-2\Delta _{0} e^{2} }{e^{2} } . \label{eq6.5a}
\end{align}
Using the Eqs. \eqref{eq6.3a} and \eqref{eq6.5a}, $A$ satisfies the following inequality:
\begin{align}
\left(-2+\frac{Ei(2) (6+\Delta _{0} )-\Delta _{0} e^{2} }{e^{2} } \right) <A \le \left(2+\frac{Ei(2) (6+\Delta _{0} )-2\Delta _{0} e^{2} }{e^{2} } \right),  \ \ \Delta _{0} \ge 0. \label{eq6.6a}
\end{align}
Differentiating \eqref{eq4.2a} with respect to $r$, we get an expression for the pressure gradient:
\begin{align}
\left(\frac{8 \pi G}{c^{4} } \frac{dp_{r} }{dr} \right)
&=2c_{0}^{2} r \Big[ A e^{2\phi } (1-4\phi +\phi ^{2} +2\phi ^{3} )+(-10+14\phi -6\phi ^{2} )\nonumber\\
&\qquad\quad \  -6 f^{2} (1+5\phi ) e^{2\phi -2} \bar{E}i (2-2\phi )\nonumber\\
&\qquad\quad \  -6 (5-12\phi -3\phi ^{2} +10\phi ^{3} )e^{2\phi -2} Ei(2-2\phi )\nonumber \\
&\qquad\quad \  +\Delta _{0} [-2 \phi +e^{2\phi -2} Ei (2-2\phi )(-2+5\phi -3\phi ^{3} )\nonumber\\
&\qquad\quad  \ -e^{2\phi -2} f^{2} (1+\phi ) \bar{E}i (2-2\phi ) ]\Big]. \label{eq6.7a}
\end{align}
Thus it is found that extremum of $p_{r}$ occurs at the centre i.e.\
\begin{align}
&p_{r}' =0 \Rightarrow r=0 \ \text{ and} \nonumber\\
&\frac{8 \pi G}{c^{4} } \left(p_{r}'' \right)_{r=0} = -\frac{2c_{0}^{2} }{e^{2} } \left[(10-A) e^{2} +6 \bar{E}i (2)+30Ei(2)+\Delta _{0}
[ 2Ei (2)+\bar{E}i (2)]\right] \label{eq6.8a}
\end{align}
This shows that the expression of right hand side of equation
\eqref{eq6.8a} is negative for all values of $A$ and $\Delta _{0}
$ satisfying condition \eqref{eq6.8a}.\ Then radial pressure
$(p_{r})$ is maximum at the centre and monotonically decreasing.

Differentiating \eqref{eq4.3a} with respect to $r$, we get
\begin{align}
\left(\frac{8 \pi G}{c^{4} } \frac{dp_{t} }{dr} \right)
&=2c_{0}^{2} r \Big[A e^{2\phi } (1-4\phi +\phi ^{2} +2\phi ^{3}
)+(-10+14\phi -6\phi ^{2} )+\Delta _{0} (1-2 \phi ) \nonumber\\
&\qquad\quad \ +e^{2\phi -2} Ei (2-2\phi ) [\Delta _{0} (-2+5\phi
-3\phi ^{3} )-6 (5-12\phi -3\phi ^{2} +10\phi ^{3} )]\nonumber\\
&\qquad\quad \ -(6+ \Delta _{0} )f^{2} (1+5\phi ) e^{2\phi -2}
\bar{E}i (2-2\phi ) \Big] \label{eq6.9a}
\end{align}
Which suggest that the extremum value of $p_{t}$ occurs at the centre i.e.
\begin{align}
&p'_{t} =0 \Rightarrow r=0 \ \text{ and}\\ &\frac{8 \pi G}{c^{4} }
\left(p''_{t} \right)_{r=0} =-\frac{2c_{0}^{2} }{e^{2} }
\left[(10-A-\Delta _{0} )e^{2} +6 \bar{E}i (2)+30Ei (2)+\Delta
_{0} [ 2Ei (2)+\bar{E}i (2)]\right]; \label{eq6.10a}
\end{align}
Under the condition \eqref{eq6.6a}, the expression of right hand
side of Eq.\,\eqref{eq6.10a} is negative for all values of $A$ and
$\Delta _{0} $.\ This shows that the tangential pressure $(p_{t})$
is maximum at the centre and monotonically decreasing.

Now differentiating equation \eqref{eq4.1a} with respect to $r$ , we get
\begin{align}
\frac{8 \pi G}{c^{2} } \frac{d\rho }{dr}
&=2c_{0}^{2} r \Big[ A e^{2\phi } (5-23\phi ^{2} +10\phi ^{3}+8\phi ^{4} ) \nonumber\\
&\qquad\quad \ + (-30-24\phi -60\phi ^{3} -48\phi ^{4} ) e^{2\phi -2} Ei(2-2\phi )\nonumber\\
&\qquad\quad \  +6 f^{2} (3-5\phi -4\phi ^{2} )e^{2\phi -2} \bar{E}i (2-2\phi )-22\phi +6\phi ^{2} \nonumber\\
&\qquad\quad \  +\Delta _{0} [e^{2\phi -2} f \{ Ei (2-2\phi ) (-5-19\phi +12\phi ^{2} +8\phi ^{3} )\nonumber \\
&\qquad\quad \  -f(-5+9\phi +8\phi ^{2} )\bar{E}i (2-2\phi )\nonumber \\
&\qquad\quad \   -2 f^{2} \phi \bar{\bar{E}}i (2-2\phi )\} -2 (-5+7\phi ) ] \Big]; \label{eq6.12a}
\end{align}
where $\bar{\bar{E}}i (2-2 \phi )=\frac{4}{(2-2\phi )^{2} } +4\sum \limits_{N=1}^{\infty }\frac{N (2-2\phi )^{N-1} }{(N+1)!} $.

The extremum of $\rho $ occur at the centre i.e.
\begin{align}
&\rho ' = 0 \Rightarrow r=0 \ \text{ and}\nonumber\\
&\frac{8 \pi G}{c^{2} } \left(\rho ''\right) _{r=0} = -\frac{2c_{0}^{2} }{e^{2} } \left[-(5A+10\Delta _{0} ) e^{2} +Ei (2)(30+5\Delta _{0} )
-\bar{E}i (2) (18+5\Delta _{0} )\right] ; \label{eq6.11a}
\end{align}
Thus, the expression of right hand side of \eqref{eq6.12a} is
negative.\ This shows that the density $\rho $ is maximum at the
centre and monotonically decreasing towards the pressure free
interface.

The square of adiabatic sound speed at the centre, $\frac{1}{c^{2} } \left(\frac{dp}{d\rho } \right)_{r=0} $, is given by
\begin{align}
&\frac{1}{c^{2} } \left(\frac{dp_{r} }{d\rho } \right)_{r=0}
=\left(\frac{(10-A) e^{2} +6 \bar{E}i (2)+30Ei(2)+\Delta _{0} [
2Ei (2)+\bar{E}i (2) ]} {(5A+10\Delta _{0} ) e^{2} +Ei (2)
(30+5\Delta _{0} )+\bar{E}i (2) (18+5\Delta _{0} )} { }\right),
\label{eq6.13a}\\ &\frac{1}{c^{2} } \left(\frac{dp_{t} }{d\rho }
\right)_{r=0} =\left(\frac{(10-A-\Delta _{0} ) e^{2} +6 \bar{E}i
(2)+30Ei(2)+\Delta _{0} [ 2Ei (2)+\bar{E}i (2) ]}{(5A+10\Delta
_{0} )e^{2} +Ei (2)(30+5\Delta _{0} )+\bar{E}i (2) (-18+5\Delta
_{0} )} \right). \label{eq6.14a}
\end{align}
The causality condition is obeyed at the centre for all values of
constants under the condition \eqref{eq6.6a} i.e. radial and
tangential velocity of sound are monotonically decreasing and less
than $1$.

\subsection*{(b) For $n=-2$}
\renewcommand{\theequation}{6.\arabic{equation}b}
\setcounter{equation}{0}

The central values of pressure and density are given by
\begin{align}
&\left(\frac{8 \pi G p_{r} }{c^{4} } \right)_{r=0}
=\left(\frac{8\pi Gp_{t} }{c^{4} } \right)_{r=0} =\frac{c_{0} }{4}
\left(4A+8+\Delta _{0} \right)\label{eq6.1b}\\[-2pt]
&\left(\frac{8 \pi G \rho }{c^{2} } \right)_{r=0} = \frac{c_{0}
}{4} \left(-12A+24+3\Delta _{0} \right)\label{eq6.2b}
\end{align}

The central values of pressure and density should be non zero positive and finite, then $A$ satisfies the following conditions::
\begin{align}
A>\left(-2-\frac{\Delta _{0} }{4} \right) \ \text{and} \  A<\left(2-\frac{\Delta _{0} }{4} \right)\label{eq6.3b}
\end{align}
Subjecting the condition that positive value of ratio of
pressure-density and should be less than $1$ at the centre i.e.
$\frac{p_{0} }{\rho _{0} c^{2} } \le 1$ which leads to the
following inequality,
\begin{align}
\left(\frac{p_{r} }{\rho c^{2} } \right)_{r=0} =\left(\frac{p_{t}
}{\rho c^{2} } \right)_{r=0} =\left(\frac{4A+8+\Delta _{0}
}{-12A+24+3\Delta _{0} } \right) \le 1\label{eq6.4b}
\end{align}
By the inequality \eqref{eq6.4b}, we get
\begin{align}
A \le \left(1+\frac{\Delta _{0} }{8} \right)\label{eq6.5b}
\end{align}
By the inequalities \eqref{eq6.3b} and \eqref{eq6.5b}, $A$ satisfies the inequality
\begin{align}
\left(-2-\frac{\Delta _{0} }{4} \right) <A \le \left(1+\frac{\Delta _{0} }{8} \right),  \  \Delta _{0} \ge 0.\label{eq6.6b}
\end{align}
Differentiating Eq.\,\eqref{eq4.2b} with respect to $r$, we get
\begin{align}
\frac{8 \pi G}{c^{4} } \frac{dp_{r} }{dr}
&=2c_{0}^{2} r \left[{\left\{-2A.+\frac{15-\Delta _{0} }{8} \log \left(\frac{1-\phi }{1+\phi } \right)\right\} f^{2} (-1+3\phi +13\phi ^{2} +9 \phi ^{3} )}\right.\nonumber \\
&\left.{-\frac{64+130x-330x^{2} -150x^{3} +270x^{4} }{8} +\frac{\Delta _{0} }{8} { (2-16}\phi +19\phi ^{2} {-10}\phi ^{3} { -3}\phi ^{4} {)}} \right]. \label{eq6.7b}
\end{align}
Thus it is found that extremum value of $p_{r}$ occur at the centre i.e.
\begin{align}
p_{r}' =0 \Rightarrow  r=0 \ \ \text{and} \ \ \frac{8 \pi G}{c^{4} } \left(p_{r}'' \right)_{r=0} =\left(\frac{ -16A-64+2\Delta _{0} }{8} \right)\label{eq6.8b}
\end{align}
Thus the expression of right hand side of the Eq.\,\eqref{eq6.8b}
is negative for all values of constants $A$ and $\Delta _{0}$
satisfying condition \eqref{eq6.6b} and it is showing that the
pressure $(p_{r})$ is maximum at the centre and monotonically
decreasing.

Differentiating Eq.\,\eqref{eq4.3b} with respect to $r$, we get
\begin{align}
\frac{8 \pi G}{c^{4} } \frac{dp_{t} }{dr}
 &=2c_{0}^{2} r \bigg[ \left(-2A+\frac{15-\Delta _{0} }{8} \log \left|\frac{1-\phi }{1+\phi } \right|\right) f^{2} (-1+3\phi +13\phi ^{2} +9 \phi ^{3} )\nonumber\\
 &\qquad\qquad -\frac{64+130x-330x^{2} -150x^{3} +270x^{4} }{8} \nonumber\\
 &\qquad\qquad +\frac{\Delta _{0} }{8} (2+43\phi ^{2} {-10} \phi ^{3} {-3}\phi ^{4} )\bigg], \label{eq6.9b}
\end{align}
Thus it is found that the extremum value of $p_{t}$ occurs at the centre
i.e.
\begin{align}
p'_{t} =0 \Rightarrow r=0 \ \ \text{and} \ \ \frac{8 \pi G}{c^{4}
} \left(p''_{t} \right)_{r=0} = 2 c_{0}^{2}
\left(\frac{16A-64+2\Delta _{0} }{8} \right), \label{eq6.10b}
\end{align}
Thus the expression of right hand side of the Eq.\,\eqref{eq6.10b}
is also negative for all values of $A$ and $\Delta _{0} $
satisfying condition \eqref{eq6.6b}. This behavior shows that
transversal pressure $(p_{t})$ is maximum at the centre and
monotonically decreasing.

Now differentiating equation \eqref{eq4.1b} with respect to $r$, the expression of density gradient as:
\begin{align}
\left(\frac{8 \pi G}{c^{2} } \frac{d\rho }{dr} \right)
&=2c_{0}^{2} r\left[ {\frac{5 S}{8 (1-\phi ^{2} )} -\left(2A+\frac{15+\Delta _{0} }{8} \log \frac{1-\phi }{1+\phi } \right)f^{2} (-5-17\phi +25\phi ^{2} +45\phi ^{3} )}\right.\nonumber \\
&\qquad\qquad\left.{-\frac{\Delta _{0} }{8} (-30+88\phi -72\phi ^{2} +2\phi ^{3} +25\phi ^{3} )} \right], \label{eq6.11b}
\end{align}
where, $S=(32+86 \phi +10 \phi ^{2} -176 \phi ^{3} -312 \phi ^{4} +90\phi ^{5} +273\phi ^{6} )$.

Thus the extremum value of $\rho $ occurs at the centre i.e.
\begin{align}
\rho ' = 0 \Rightarrow r=0 \ \text{and} \  \frac{8 \pi G}{c^{2} } \left(\rho ''\right) _{r=0} =2c_{0}^{2} \left(\frac{ 180+80A+30\Delta _{0} }{8} \right)>0 ; \label{eq6.12b}
\end{align}
The right hand side of Eq.\,\eqref{eq6.12b} is showing positive
due to the inequality \eqref{eq6.6b}, and this condition gives
that the energy density $\rho $ is minimum at centre and
monotonically increasing.

and the square of its adiabatic sound speed at the centre, $\frac{1}{c^{2} } \left(\frac{dp}{d\rho } \right)_{r=0} $, is given by
\begin{align}
\frac{1}{c^{2} } \left(\frac{dp_{r} }{d\rho } \right)_{r=0}
=\frac{1}{c^{2} } \left(\frac{dp_{t} }{d\rho } \right)_{r=0}
=\frac{\left(16A-64+3\Delta _{0} \right)}{ \left[180+80A+30\Delta
_{0} \right]} \label{eq6.13a}
\end{align}

The causality condition is negative at the centre for all values
of constants satisfying condition \eqref{eq6.6b}. Due to
increasing nature of energy density, the solution is not well
behaved for $n=-2$.

\subsection*{Case~3: $n=-3$}
\renewcommand{\theequation}{6.\arabic{equation}c}
\setcounter{equation}{0}

 The central values of pressure and density are given by
\begin{align}
&\left(\frac{8 \pi G p_{r} }{c^{4} } \right)_{r=0}
=\left(\frac{8\pi Gp_{t} }{c^{4} } \right)_{r=0} =c_{0}
\left[A+356+\Delta _{0} \right]\label{eq6.1c}\\ &\left(\frac{8 \pi
G \rho }{c^{2} } \right)_{r=0} = \frac{c_{0} }{81}
\left[-243A+21320-243\Delta _{0} \right]\label{eq6.2c}
\end{align}

The central values of pressure and density should be non zero
positive and finite. Then $A$ satisfies the following conditions:
\begin{align}
A>\left(-356-\Delta _{0} \right) \ \ \text{ and } \ \ A <\left(\frac{21320}{243} -\Delta _{0} \right)\label{eq6.3c}
\end{align}

Subject to the condition that the ratio of pressure-density is
positive and less than $1$ at the centre i.e. $\frac{p_{0} }{\rho
_{0} c^{2} } \le 1$ which leads to the following inequality,
\begin{align}
\left(\frac{p_{r} }{\rho c^{2} } \right)_{r=0} =\left(\frac{p_{t}
}{\rho c^{2} } \right)_{r=0} =\left(\frac{81 \left[A+356+\Delta
_{0} \right]}{-243A+21320-243\Delta _{0} } \right)\le 1
\label{eq6.4c}
\end{align}
Eq.\,\eqref{eq6.4c} leads to
\begin{align}
A \le -\left(\frac{7516}{324} +\Delta _{0} \right)\label{eq6.5c}
\end{align}

By using the Eqs. \eqref{eq6.3c} and \eqref{eq6.5c}, we get the inequality for $A$ as:
\begin{align}
-\left(356+\Delta _{0} \right) <A \le - \left(\frac{7516}{324} +\Delta _{0} \right)\label{eq6.6c}
\end{align}

Differentiating \eqref{eq4.2c} with respect to $r$, we get
\begin{align}
\frac{8 \pi G}{c^{2} } \frac{dp_{r} }{dr}
&=2c_{0}^{2} r \left[{A { p}_{{1}} (\phi )-(411-292\phi -11460\phi ^{2} -6720\phi ^{3} +8000\phi ^{4} )+\frac{320}{243} f^{3} (1+5\phi )} \right.\nonumber\\ &\qquad\quad \ \left.{+\frac{320}{243} {p}_{{2}} (\phi )\log \frac{1+2\phi }{1-\phi } +\Delta _{{0}} {(2-30}\phi +33\phi ^{2} +{1}60\phi ^{3} { -345x}^{{4}} +{1}80 \phi ^{5} {)}} \right],\label{eq6.7c}
\end{align}
where,
\begin{align*}
{p}_{{1}} (\phi )&={(3-24}\phi {-6}\phi ^{{2}} +{132}\phi ^{{3}} {-165}\phi ^{{4}} +{60}\phi ^{{5}} {)}, \\
{p}_{{2}} (\phi )&=(15-40 \phi -252 \phi ^{2} -336 \phi ^{3} +120 \phi ^{4} )
\end{align*}
Thus it is found that extremum value of $p_{r}$ occurs at the centre i.e.
\begin{align}
p_{r}' =0 \Rightarrow r=0 \ \text{ and } \ \frac{8 \pi G}{c^{4} } \left(p_{r}'' \right)_{r=0} =2c_{0}^{2} \left(3A-411+\frac{320}{243} +2\Delta _{0} \right). \label{eq6.8c}
\end{align}

So the expression of right hand side of Eq.\,\eqref{eq6.8c} is
negative for all values of $A$  and $\Delta _{0} $  satisfying the
condition \eqref{eq6.6c}. This shows that the pressure $(p_{r})$
is maximum at the centre and monotonically decreasing.

Differentiating \eqref{eq4.3c} with respect to $r$, we get
\begin{align}
\left(\frac{8 \pi G}{c^{2} } \frac{dp_{t} }{dr} \right)
&=2c_{0}^{2} r \bigg[ {A { (3-24}\phi {-6}\phi ^{{2}} +{132}\phi
^{{3}} {-165}\phi ^{{4}} +{60}\phi ^{{5}} {)}} \nonumber\\
&\qquad\quad \ \ {-(411-292\phi -11460\phi ^{2} -6720\phi ^{3}
+8000\phi ^{4} )+\frac{320}{243} f^{3} (1+5\phi )} \nonumber\\
&\qquad\quad \ \  {+\frac{320}{243} (15-40 \phi -252 \phi ^{2}
-336 \phi ^{3} +120 \phi ^{4} ) \log \frac{1+2\phi }{1-\phi } }
\nonumber\\ &\qquad\quad \ \  {+\Delta _{{0}} {(2-30}\phi +36\phi
^{2} +{176}\phi ^{3} {-325x}^{{4}} +{1}80 \phi ^{5} {)}}\bigg].
\label{eq6.9c}
\end{align}

Thus it is found that extrema of $p_{t}$ occurs at the centre i.e.
\begin{align}
p'_{t} =0 \Rightarrow r=0  \ \ \text{and} \frac{8 \pi G}{c^{4} }
\left(p''_{t} \right)_{r=0} =2c_{0}^{2}
\left(3A-411+\frac{320}{243} +2\Delta _{0} \right).\label{eq6.10c}
\end{align}

Thus it is clear that the expression of right hand side of
Eq.\,\eqref{eq6.10c} is negative for all values of $A$ and $\Delta
_{0} $ satisfying the condition \eqref{eq6.6c}. This condition
gives that the transversal pressure is maximum at the centre and
monotonically decreasing.

Now differentiating equation \eqref{eq4.1c} with respect to r we get
\begin{align}
\left(\frac{8 \pi G}{c^{2} } \frac{d\rho }{d\phi } \right)
&=2c_{0}^{2} r\bigg[{A (15-270x^{2} +660x^{3} -585x^{4} +204x^{5} )+\frac{\rho _{1} }{81} } \nonumber\\
&\qquad\quad \ \  {+(9-140x+540 x^{3} -495x^{4} +156x^{5} )\left\{\frac{320}{81} \log \frac{1+2\phi }{1-\phi } \right\}} \nonumber\\
&\qquad\quad \ \  {+{ }\Delta _{{0}} {(2-30}\phi +33x^{2} +{160}\phi ^{3} { -345x}^{{4}} +{180}\phi ^{5} {) }}\bigg]\label{eq6.11c}
\end{align}
\[
\rho _{1} =(-26885-55018 \phi +64686 \phi ^{2} -1149772 \phi ^{3} +685612 \phi ^{4} -296960 \phi ^{5} +128960 \phi ^{6} )
\]
Thus the extrema of $\rho $ occur at the centre i.e.
\begin{align}
\rho ' = 0 \Rightarrow r=0 \ \text{ and } \ \frac{8 \pi G}{c^{2} } \left(\rho ''\right) _{r=0} =2c_{0}^{2} \left(\frac{-26885}{81} +15A+2\Delta _{0} \right)\label{eq6.12c}
\end{align}

Thus, the expression of right hand side of Eq.\,\eqref{eq6.12c} is
negative for all values of $A$ and $\Delta _{0}$ satisfying the
condition \eqref{eq6.6c}.Then density $\rho $ is maximum at the
centre and monotonically decreasing.

The square of adiabatic sound speed at the centre, $\frac{1}{c^{2} } \left(\frac{dp}{d\rho } \right)_{r=0} $, are given by
\begin{align}
\frac{1}{c^{2} } \left(\frac{dp_{r} }{d\rho } \right)_{r=0}
=\frac{1}{c^{2} } \left(\frac{dp_{t} }{d\rho } \right)_{r=0}
=\left(\frac{-99553-729A+486\Delta _{0} }{3
[-26885+1215A+162\Delta _{0} ]} \right)\label{eq6.13c}
\end{align}

The causality condition is less than 1 and positive at the centre
for all values of constants for all values of $A$ and $\Delta _{0}
$ satisfying condition \eqref{eq6.6c}.

\section{Boundary conditions for evaluation of constants $A$, $D$ and $c_0$}\label{s7}
\renewcommand{\theequation}{7.\arabic{equation}}
\setcounter{equation}{0}

The above system of equations is to be solved subject to the
boundary condition that radial pressure $p_{r} =0$ at $r=a$
(where, $r=a$ is the outer boundary of the fluid sphere). It is
clear that $m (r=a)=M$ is a constant and, in fact, the interior
metric \eqref{eq2.1} can be joined smoothly at the surface of
spheres $(r=a)$, to an exterior Schwarzschild metric whose mass is
the same as in above i.e. $m(r=a)=M$ (Masiner and Sharp
\cite{45}).

Then the interior metric of this fluid spheres should be joined
smoothly with Schwarzschild exterior metric such as
$B^{2}(a)=1-2u$, where $u=M/a$, where $M$ is the mass of the fluid
sphere as measured by the exterior field and $a$ is the boundary
of the sphere.

By joining \eqref{eq3.5a} and \eqref{eq3.5c} on the boundary of
the anisotropic fluid spheres $(r=a)$ and by setting $\phi _{a}
=c_{o} a^{2} $ and $\psi _{anis} (a)=1-2 u_{anis} $, we get the
expressions of mass for $n=-1$ and $-3$ as:
\begin{align}
M_{anis}
&=\frac{a}{2} \phi _{a} \bigg[-A f_{a}^{3} e^{2 \phi _{a} } +(4 \phi _{a} -3\phi _{a} ^{2} +2\phi _{a} ^{3} )\nonumber\\
&\qquad\qquad +(6 +\Delta _{0} \phi _{a} )f_{a}^{3} e^{2\phi _{a} -2} Ei (2-2 \phi _{a} )-\Delta _{0} \phi _{a} f_{a}^{2} \bigg],\label{eq7.1}\\
M_{anis}
&=\frac{a}{2} \phi _{a} \bigg[\frac{M_{1} }{81} -A(1-\phi _{a} )^{5} (1+2\phi _{a} )\nonumber\\
&\qquad \qquad +\frac{320}{243} (1-\phi _{a} )^{5} (1+2\phi _{a} )\log \frac{1+2\phi _{a} }{1-\phi _{a} } -M_{2} \Delta _{0} \phi _{a} . f_{a}^{3} \bigg],\label{eq7.2}\\
M_{1} &=(292-305\phi _{a} -662\phi _{a} ^{2} +1796 \phi _{a} ^{3} -1360 \phi _{a} ^{4} +320 \phi _{a} ^{5} ), \nonumber\\
M_{2} &=(1+2\phi _{a} ) (1-3\phi _{a} +3\phi _{a}^{2} ),\nonumber\\
S&=(16+17\phi _{a} -50 \phi _{a} ^{2} +10 \phi _{a} ^{3} +30 \phi _{a} ^{4} -15 \phi _{a} ^{5} ).\nonumber
\end{align}
The arbitrary constant $A$ in the expression \eqref{eq3.5a} and
\eqref{eq3.5c} can be determined by using the radial pressure
$p_{r}$ is zero at the boundary, the expressions of constant $A$
for $n=-1$ and $-3$ are given as:
\begin{align} \label{eq7.3}
A&=\frac{[6(1+5\phi _{a} )+\Delta _{0} f_{a} ] f_{a}^{2} e^{2\phi _{a} -2} Ei(2-2\phi _{a} )-(2-10\phi +7\phi ^{2} -2\phi ^{3} )-\Delta _{0} f_{a}^{2} }{f_{a}^{2} (1+\phi _{a} ) e^{2\phi _{a} } } ;\\
\label{eq7.4}
A &=\frac{\left[\begin{array}{l}\!\! {(-356+411\phi _{a} +146\phi _{a} ^{2} -3820 \phi _{a} ^{3} -1680 \phi _{a} ^{4} +1600 \phi _{a} ^{5} )} \\[8pt] { -\dfrac{320}{243} f_{a}^{ 4} (1+2\phi _{a} )(1+5\phi _{a} )\log \dfrac{1+2\phi _{a} }{1-\phi _{a} } -\Delta _{0} f_{a}^{2} (1+4\phi -8\phi ^{2} -9\phi ^{3} +30\phi ^{4} )} \!\!\end{array}\right]}{f_{a}^{ 4} (1+7\phi _{a} +10\phi _{a} ^{2} )} \,.
\end{align}

Also the arbitrary constants $D$ in the metric potential for the
case $n=-1$ and $-3$ can be computed by the condition
$B(a)=1-2u_{anis} $,
\begin{align} \label{eq7.5}
D&=f_{a}^{3} [A \phi _{a} f_{a} e^{2\phi _{a} } +(1-2\phi _{a} ) -(6 +\Delta _{0} )f_{a} \phi _{a} e^{2\phi _{a} -2} Ei (2-2\phi _{a} )+ \Delta _{0} \phi _{a} ];\\
D
&=f_{a}^{6} \bigg[{A \phi _{a}  f_{a}^{2} (1+2\phi _{a} )-\frac{1}{243} f_{a}^{2} \left[\frac{3 g}{f_{a}^{3} } -320 \phi _{a} \log \left(\frac{1+2\phi _{a} }{f_{a} } \right) (1+2\phi _{a} )\right]}\nonumber \\
&\qquad\qquad {+\Delta _{0} \phi _{a}  (1+2\phi _{a} )(1-3\phi _{a} +3\phi _{a}^{2} )}\bigg].\label{eq7.6}
\end{align}
The positive constant $c_{0}$ can be calculated by taking the
surface density $2\times 10^{14}$ gm/cm$^{3} $ and using the
condition $c_{0} =\left(\frac{8\pi G}{c^{2} } a^{2} \rho _{a}
\right)/\left(1-\psi _{a} -a \psi '_{a} \right)$ for $n=-1$ and
$-3$.

\begin{figure}[h]
\centering
\includegraphics[width=5cm]{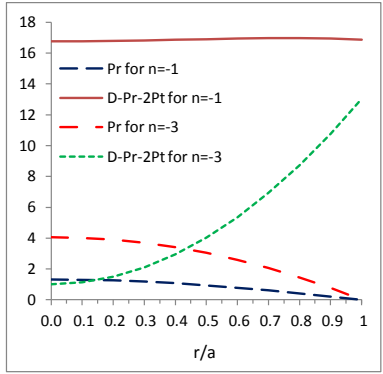}
\caption{Variations of radial pressure $P_r$ and the trace $D -
P_{r} -2 P_{t}$ of the energy-momentum tensor for $n=-1$, $c_{0}
a^{2}=0.0829$, $\Delta_{0}=4.2395$ and $n=-3$,
$c_{0}a^{2}=0.0028$, $\Delta_{0}=0.7094$}
\end{figure}

\subsection*{Tables for Numerical Values of physical quantities}

In Tables 1-2: $Z=\,$ red shift, Solar mass $M_{\Theta } =$ 1.475
km, $G=6.673\times 10^{-8}$cm$^{3}$/gs$^{2}$, $c=2.997\times
10^{10}$ cm/s, $D=(8\pi G/c^{2} c_{0} )\rho $, $P_{r} =(8\pi
G/c^{4} c_{0} )p_{r} a^{2} $, $P_{t} =(8\pi G/c^{4} c_{0} )p_{t}
a^{2} $, $\gamma =\frac{p+c^{2} \rho }{p} \frac{dp}{c^{2} d\rho }
$.

\begin{figure}[h]
\centering
\includegraphics[width=5cm]{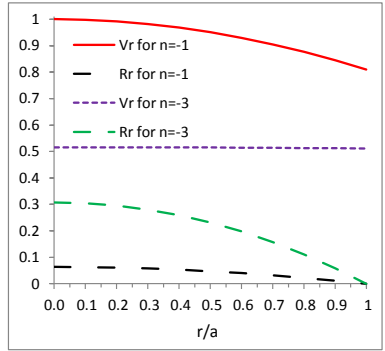}
\caption{Variations of the radial velocity $V_r=\sqrt{\frac{dp_{r}
}{c^{2} d\rho } } $ and ratio of radial pressure and density
$R_r=\frac{p_{r} }{c^{2} \rho } $ of the energy-momentum tensor
for $n=-1$, $c_{0}a^{2}=0.0829$, $\Delta_{0}=4.2395$ and $n=-3$,
$c_{0}a^{2}=0.0028$, $\Delta_{0}=0.7094$.}\label{f2}
\end{figure}

\begin{figure}[h]
\centering
\includegraphics[width=5cm]{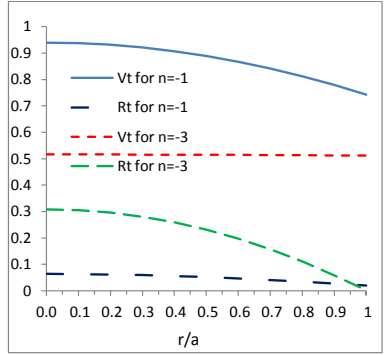}
\caption{Variations of the tangential velocity
$V_t=\sqrt{\frac{dp_{t} }{c^{2} d\rho } } $ and ratio of
tangential pressure and density $R_t=\frac{p_{t} }{c^{2} \rho } $
of the energy-momentum tensor for $n=-1$, $c_{0}a^{2}=0.0829$,
$\Delta_{0}=4.2395$ and $n=-3$, $c_{0}a^{2}=0.0028$,
$\Delta_{0}=0.7094$.}\label{f3}
\end{figure}

\begin{figure}[h]
\centering
\includegraphics[width=5cm]{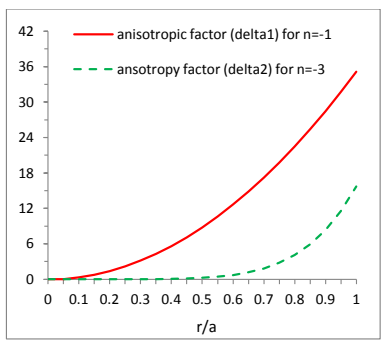}
\caption{Variations of the anisotropy factor $\Delta_1= \Delta
\times 10^{-2}$ and anisotropy factor $\Delta_2= \Delta \times
10^{-9}$ of the energy-momentum tensor for $n=-1$,
$c_{0}a^{2}=0.0829$, $\Delta_{0}=4.2395$ and $n=-3$,
$c_{0}a^{2}=0.0028$, $\Delta_{0}=0.7094$.}\label{f4}
\end{figure}

\begin{table}[H]\centering
\caption{$n$=-1, $\Delta _{0} $= 4.2395, $c_{0} a^{2} $=0.0829, Radius ($a$) = 16.0780 Km, Mass ($M$) = 1.7609 $M_{\Theta } $}\label{t1}

\setlength{\tabcolsep}{1mm}
\renewcommand{\arraystretch}{1.2}
\begin{tabular}{|c|c|c|c|c|c|c|c|c|c|} \hline
$r/a$& $P_{r}$ & $P_{t}$ & $D$ &$\Delta$ & $\sqrt{\frac{dp_{r}
}{c^{2} d\rho } } $ & $\sqrt{\frac{dp_{t} }{c^{2} d\rho } } $ &
$\frac{p_{r} }{c^{2} \rho } $ & $\frac{p_{t} }{c^{2} \rho } $
&$Z$\\ \hline 0.0 & 1.3165 & 1.3165 & 20.7045 & 0.0000 & 0.9999 &
0.9394 & 0.0636 & 0.0636 & 0.2570 \\ \hline 0.2 & 1.2543 & 1.2683
& 20.5763 & 0.0140 & 0.9918 & 0.9312 & 0.0610 & 0.0616 & 0.2554 \\
\hline 0.4 & 1.0723 & 1.1285 & 20.1933 & 0.0562 & 0.9680 & 0.9070
& 0.0531 & 0.0559 & 0.2504 \\ \hline 0.6 & 0.7852 & 0.9117 &
19.5603 & 0.1265 & 0.9292 & 0.8672 & 0.0401 & 0.0466 & 0.2421 \\
\hline 0.8 & 0.4167 & 0.6417 & 18.6860 & 0.2249 & 0.8761 & 0.8125
& 0.0223 & 0.0343 & 0.2303 \\ \hline 1.0 & 0.0000 & 0.3515 &
17.5839 & 0.3515 & 0.8092 & 0.7427 & 0.0000 & 0.0200 & 0.2151 \\
\hline
\end{tabular}

\end{table}

\begin{figure}[h]
\centering
\includegraphics[width=5cm]{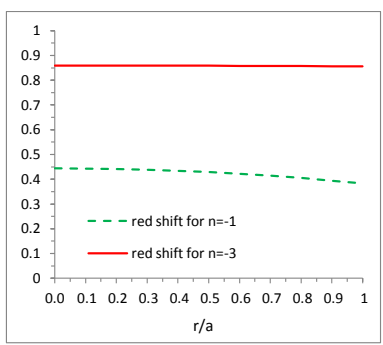}
\caption{Variations of the red-shift $(Z)$ of the energy-momentum
tensor for $n=-1$, $c_{0}a^{2}=0.0829$, $\Delta_{0}=4.2395$ and
$n=-3$, $c_{0}a^{2}=0.0028$, $\Delta_{0}=0.7094$.}\label{f5}
\end{figure}

\begin{table}[H]\centering
\caption{$n$=-3, $\Delta _{0} $=0 .7094, $c_{0} a^{2} $=0.0028, Radius ($a$) =3.1274Km, Mass ($M$) =0.8672$M_{\Theta } $}\label{t2}

\setlength{\tabcolsep}{1mm}
\renewcommand{\arraystretch}{1.2}
\begin{tabular}{|c|c|c|c|c|c|c|c|c|c|} \hline
$\frac{r}{a} $ & $P_r$ & $P_{t}$ & $D$ & $\Delta \times 10^{13} $
& $\sqrt{\frac{dp_{r} }{c^{2} d\rho } } $ & $\sqrt{\frac{dp_{t}
}{c^{2} d\rho } } $ & $\frac{p_{r} }{c^{2} \rho } $ & $\frac{p_{t}
}{c^{2} \rho } $ &$Z$\\ \hline 0.0 & 4.0619 & 4.0619 & 13.1901 & 0
& 0.5160 & 0.5160 & 0.3079 & 0.3079 & 0.8597 \\ \hline 0.2 &
3.8982 & 3.8982 & 13.1838 & 0.00997 & 0.5158 & 0.5158 & 0.2957 &
0.2957 & 0.8595 \\ \hline 0.4 & 3.4077 & 3.4077 & 13.1650 &
0.63900 & 0.5153 & 0.5153 & 0.2588 & 0.2588 & 0.8590 \\ \hline 0.6
& 2.5922 & 2.5922 & 13.1336 & 7.2949 & 0.5144 & 0.5144 & 0.1974 &
0.1974 & 0.8582 \\ \hline 0.8 & 1.4549 & 1.4549 & 13.0896 & 41.116
& 0.5132 & 0.5132 & 0.1111 & 0.1111 & 0.8571 \\ \hline 1.0 &
0.0000 & 1.5748*$10^{-8} $ & 13.0331 & 157.48 & 0.5116 & 0.5116 &
0.0000 & 1.208*$10^{-9} $ & 0.8556 \\ \hline
\end{tabular}
\end{table}

\section{Stability of the stellar structure}\label{s8}

For physically acceptable model, one aspect that the velocity of
sound should be within the range $0\le v_{a}^{2} =(dp/c^{2} d\rho
)\le 1$ (Herrera \cite{47} and Abreu et al.\ \cite{48}).\ In
present models, the expression for velocity of sound at the centre
is given by equations (6.13a), (6.14a) and (6.13c). We plot the
radial and transverse velocity of sound in Fig.~3 and conclude
that all parameters satisfy the inequalities $0=v_{a r}^{2}
=(dp_{r} /c^{2} d\rho ) \leq 1$ and $0=v_{a t}^{2} =(dp_{t} /c^{2}
d\rho ) \leq 1$, everywhere inside the star models. From equations
(6.13a), (6.14a) and (6.13c), we found that $|v_{a t}^{2} -v_{a
r}^{2} |\le 1$ at the centre and proved that velocity of sound in
monotonically decreasing throughout inside the star. Also $0\le
v_{a t}^{2} \le 1$ and $0\le v_{a r}^{2} \le 1$, therefore
$\left|v_{a t}^{2} -v_{a r}^{2} \right|\le 1$. Now, to examine the
stability of local anisotropic fluid distribution, Herrera's
\cite{46} proposed the cracking (also known as overturning)
concept which states that the region, in which radial speed of the
sound is greater than transverse speed of the sound, is a
potentially stable region.

In our proposed models, the models are stable with the radius
16.0780 Km, $\Delta _{0} = 4.2395$, $c_{0} a^{2} =0.0829 $ for
$n=-1$ and radius 3.1274Km, $\Delta _{0} =0 .7094$, $c_{0} a^{2}
=0.0028$ for $n=-3$.

\section{Physical analysis and conclusions}\label{s9}

In the present paper, the new set of anisotropic exact solution of
Einstein's field equations we have presented by taking the metric
potential $g_{44}=(1-c_{0}r^{2})^{n}$ for $n=-1, -2$ and $-3$ and
specific choice of anisotropic factor $\Delta $ which involves the
anisotropic parameter $\Delta_{0}$. The obtained solutions are
utilized to contract the super dense star models with surface
density $2\times 10^{14}$ gm/cm$^{3}$. It is observed that
solutions are satisfying all reality and physical conditions
(mention its in Section~\ref{s5}) for $n=-1$ and $-3$. But the
solution is not compatible for $n=-2$ due to increasing nature of
its density (Section~\ref{s6}, Case (b)). The anisotropic fluid
sphere possesses the maximum mass and corresponding radius are
$1.7609M_{\Theta }$ and 16.0780 Km for $n=-1$,
$c_{0}a^{2}=0.0829$, $\Delta_{0}=4.2395$ and $0.8672 M_{\Theta }$
and 3.1274Km for $n=-3$, $c_{0}a^{2}=0.0028$, $\Delta_{0}=0.7094$.
The red shift for $n=-1$ and $-3$ are monotonically decreasing
towards the pressure free interface $r=a$ and found that the red
shift at the centre $(Z_{0})$ and at surface $(Z_{a})$ are: (i)
$Z_{0}=0.2570$ and $Z_{a}=0.2151$ for $n=-1$, (ii) $Z_{0}=0.8597$
and $Z_{a}=0.8556$ for $n=-3$ for both strong energy and dominated
energy conditions. In our models, the red shift is also satisfied
the upper bound limit for the realistic anisotropic star models
(Ivanov \cite{4}) and its behavior is represented by the
Fig.~\ref{s5}. The Tables~\ref{t1} and \ref{t2} shows the
numerical values of physical parameters. Fig. 1 shows that the
fluid spheres satisfies the strong energy condition. The behaviors
of velocity and pressure density ratio are given by the
Fig.~\ref{f2} and \ref{f3}. Fig.~\ref{f4} represents the
increasing nature of anisotropy factor for the fluids spheres.

\section*{Acknowledgements}

The authors are very grateful to the Honorable Editors and
Referees for their valuable comments and suggestions, which made
the paper in a more presentable form and also grateful to the
University of Nizwa, Sultanate of Oman, for providing all the
necessary facilities and encouragement.

\end{document}